# Scalable and ultralow power silicon photonic two-dimensional phased array

**Michelle Chalupnik[1,2,5]\*, Anshuman Singh[1], James Leatham[3], Marko Loncar[4], Moe Soltani[1]**

[1]*Raytheon BBN, 10 Moulton Street Cambridge, MA 02138, USA*
[2]*Department of Physics, Harvard University, Cambridge, MA 02138, USA*
[3]*Raytheon Intelligence & Space, 2000 El Segundo Dr., El Segundo, CA, USA*
[4]*John Paulson School of Engineering and Applied Science, Harvard University, Cambridge, MA 02138, USA*
[5]*Currently with Aliro Quantum, Brighton, MA 02135, USA*

*michelle.chalupnik@raytheon.com, mo.soltani@raytheon.com*

**Abstract:** Photonic integrated circuit based optical phased arrays (PIC-OPA) are emerging as promising programmable processors and spatial light modulators, combining the best of planar and free-space optics. Their implementation in silicon photonic platforms has been especially fruitful. Despite much progress in this field, demonstrating steerable two-dimensional (2D) OPAs scalable to a large number of array elements and operating with a single wavelength has proven a challenge. In addition, the phase shifters used in the array for programming the far field beam are either power hungry or have a large footprint, preventing implementation of large scale 2D arrays. Here, we demonstrate a two-dimensional silicon photonic phased array with high-speed (~330 KHz) and ultralow power microresonator phase-shifters with a compact radius (~3 μm) and $2\pi$ phase shift ability. Each phase-shifter consumes an average ~250 μW static power for resonance alignment and ~50 μW power for far field beamforming. Such PIC-OPA devices can enable a new generation of compact and scalable low power processors and sensors.

# Introduction

Integrated optical phased arrays (OPA) in a silicon photonic integrated circuit (PIC) platform are a form of on-chip programmable spatial light modulators (SLM) and have shown great promise for a broad range of applications including holography, light detection and ranging (LiDAR), and free-space communications [1-4]. PIC-OPAs can enable formation of complex beam patterns as well as precise and high speed beam steering for LiDAR or sensing. As compact and high-speed programmable SLMs, PIC-OPAs promise a new generation of advanced on-chip optical processor engines [5-7] which can replace existing devices and their associated bulky components [8].

Despite significant advancements in the PIC-OPA technology, most demonstrated works rely on waveguide phase shifters which are either power hungry or have large footprints [9-16], inhibiting the scalability of two-dimensional (2D) phased arrays to larger numbers of array elements. The majority of 2D beamforming and beam steering work using PICs relies on tuning the input laser for steering in one direction and programming the phase shifters for steering in the other orthogonal direction [e.g. 3, 10, 12, 16, 17]. Some other 2D phased array work places phase shifters outside the array, however these approaches do not show scalability due to their optical routing topology limitations [11, 15], and experience larger optical insertion loss for larger arrays [15]. 2D phased array based on MEMS [14] have been demonstrated with promising results. However, the required large voltage (~10 V) for the MEMS components and the custom nature of device fabrication make these devices incompatible with CMOS driving circuits as well as CMOS manufacturing. The recent interesting work in [6] designs and employs an array of silicon photonic crystal (PC) cavities for making a 2D SLM wherein each PC cavity acts simultaneously as a phase shifter and an antenna [6]. However, each cavity which is tuned optically with a different wavelength provides a maximum phase shift of $0.2\pi$ limiting the full range of phase programming. In addition, the entire PC layer is suspended which limits the maximum power that each cavity can handle before transitioning to thermal and absorption nonlinearity regimes.

Recently silicon ring resonator phase shifters have proven to be a solution for small footprint and low power consumption phase shifters in optical phased arrays, allowing a path for phased array scalability [18]. When these resonators are designed in the strong overcoupling regime, they can provide a phase shift of 2π for their corresponding antenna with low power consumption. Optical phased arrays with silicon ring resonator phase shifters are also relatively robust and can be fabricated reproducibly in standard CMOS foundries.

Here, we present a two-dimensional scalable OPA in a Si PIC platform wherein each individual antenna element has its own ultralow power phase shifter consisting of an overcoupled ring resonator. The resonator phase shifters in this work are tuned using Si-doped microheaters integrated in close proximity to the resonators. With an 8×8 OPA with overcoupled ring resonator phase shifters, we demonstrate beamforming and beam steering, verifying the low power consumption required for each application, and showing promise of this technology as a compact and low power on-chip SLM.

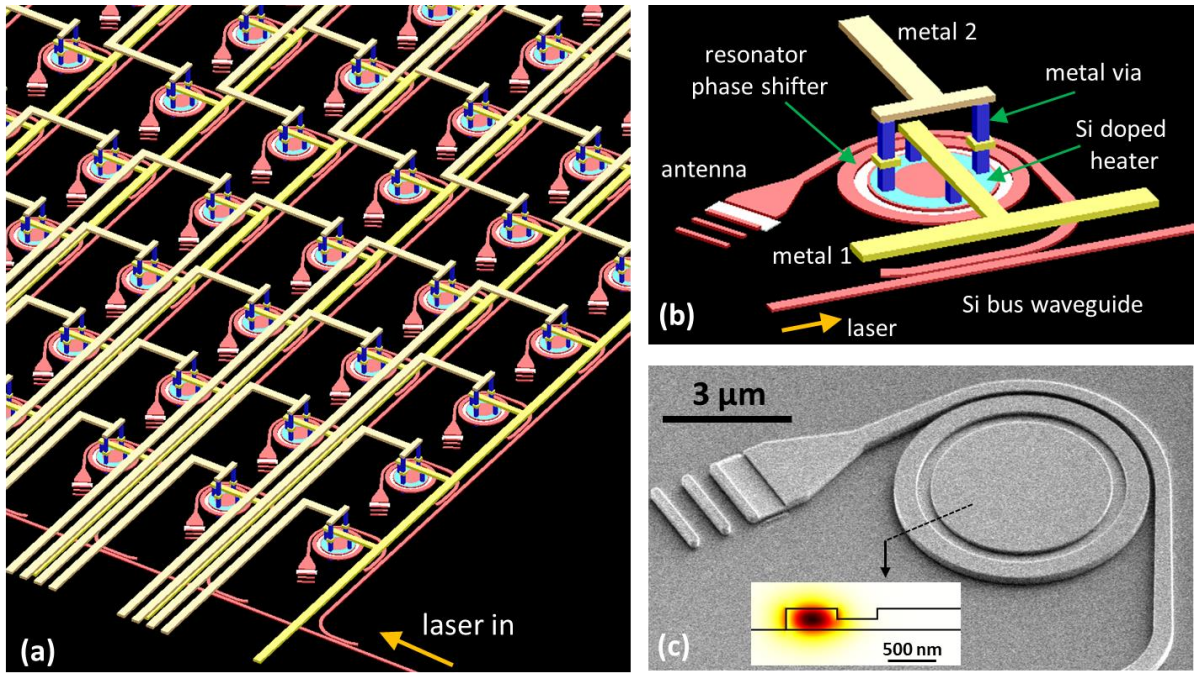

**Fig. 1**: **PIC-OPA structure overview**: (a) A schematic view of a portion of the 8 x 8 PIC-OPA designed for this work. Light enters the array through a waveguide, and using tap coupler is split into eight bus waveguides with equal power. For each bus waveguide there are eight tap couplers to extract the light and pass to a ring resonator phase shifter before feeding into an antenna. All tap couplers are designed such that the antenna elements receive equal intensities. (b) Schematic of a unit cell of the PIC OPA. Laser light is tapped from a Si waveguide, goes through a strongly overcoupled ring resonator phase shifter (~3 µm external radius), and then is sent to the antenna for radiation. The interior of the ring resonator is doped to create a resistive heater, and has been segmented into four parallel resistor segments to reduce the overall resistance. Metal lines in two different layers connect these resistor segments and make them parallel. (c) A scanning electron microscope view of the unit cell before adding the metal contacts. The inset in (c) shows the cross section transverse-electric (TE) mode profile of the ring resonator.

**PIC-OPA Design**

Our PIC-OPA device is designed for large two-dimensional scaling with minimal power consumption. Fig. 1(a) shows the schematic of a portion of the PIC-OPA architecture, containing many unit cells and displaying the optical routing. Laser light enters the array through a waveguide, and using tap couplers is split into eight bus waveguides with equal optical power. For each bus waveguide there are eight OPA unit cells consisting of tap couplers to extract the light as input to each OPA unit cell which is shown in Fig. 1(b). As shown in this figure, the laser signal after the tapped coupler is sent to a ring resonator phase shifter before feeding the antenna of the unit cell. The antenna design is a grating-based and optimized for maximum radiation efficiency [18]. All tap couplers are designed such that the antenna elements receive equal intensities, a necessary design consideration to prevent brightness inhomogeneities in the far field when beamforming and beam steering. The optical routing within the array is scalable to realize a 2D OPA with a large number of antenna elements [1]. An SEM of one unit cell on the device with the overcoupled ring resonator and antenna is shown in Fig. 1(c).

The ultralow power phase shifter is realized by Si-doped microheaters implemented in the interior region of the resonator as shown in Fig. 1(b). The doped region, which is n-type, is segmented into four parallel resistor segments to reduce the overall resistance of the heater to sub-k$\Omega$ values. Two layers of metal contacts as shown in Fig. 1b make these resistor segments parallel. A partially etched region separates the ring resonator region from the doped region. The purpose of this separation is to further confine the optical mode in the ring (see the inset of Fig. 1(c)) and minimize the optical absorption due to the doping region which can degrade the resonator's intrinsic Q. The remaining silicon layer in the partially etched region allows efficient heat transfer from the heater to the optical mode of the ring [19].

In this work, for the proof of concept, we demonstrate an 8×8 OPA with 64 unit cells. Detailed design parameters are included in the Supplementary Materials. In current array fabrication, the metal routing to all the phase shifters lies in the same optical layer (see Fig. 1(b)) limiting the array size. However, the architecture scales to larger sizes when electronic addressing is integrated under the optical layer [20].

To enable a ring resonator to operate as a phase shifter, the coupling between the resonator and the external waveguide needs to be in the strong overcoupling regime. The transmission from a ring resonator can be derived as [18]

$$T = \frac{2i(\omega - \omega_0)/\omega_0 + 1/Q_0 - 1/Q_C}{2i(\omega - \omega_0)/\omega_0 + 1/Q_0 + 1/Q_C}, \qquad Eq.1$$

where $\omega$ is the laser frequency, $\omega_0$ is the resonant frequency of the resonator, $Q_0$ is the intrinsic quality factor of the resonator, and $Q_C$ is the waveguide-resonator coupling quality factor. In the overcoupled regime, $Q_c/Q_0 \ll 1$, and in this case, the resonator phase is dependent on the coupling quality factor and is approximately equal to [18]

$$\phi \approx -2 \tan^{-1} \frac{2\, Q_C(\omega - \omega_0)}{\omega_0}. \qquad Eq.2$$

Figure 2 shows the transmission intensity and the phase spectrum of a ring resonator coupled to a waveguide for different coupling regimes. As shown in Fig. 2(b), in the strong overcoupling regime ($Q_c/Q_0 \ll 1$) we achieve close to $2\pi$ phase variation when sweeping the resonance and the laser wavelength relative to each other. Additionally, as shown in Fig. 2(a), the amplitude modulation for transmission in the overcoupled regime is minimal compared to the undercoupled regime, where $Q_c/Q_0 >> 1$, preventing distortions in brightness during the phase shifting process. Therefore, to achieve a strong overcoupling, careful design of the waveguide-resonator coupling is required. As shown in Fig. 1b, we use a pulley coupling scheme to create a long interaction length between the waveguide and the resonator. In addition, the waveguide width and the gap between the waveguide and the resonator are designed to provide maximum phase matching and efficient modal overlap between the waveguide mode and the resonator mode.

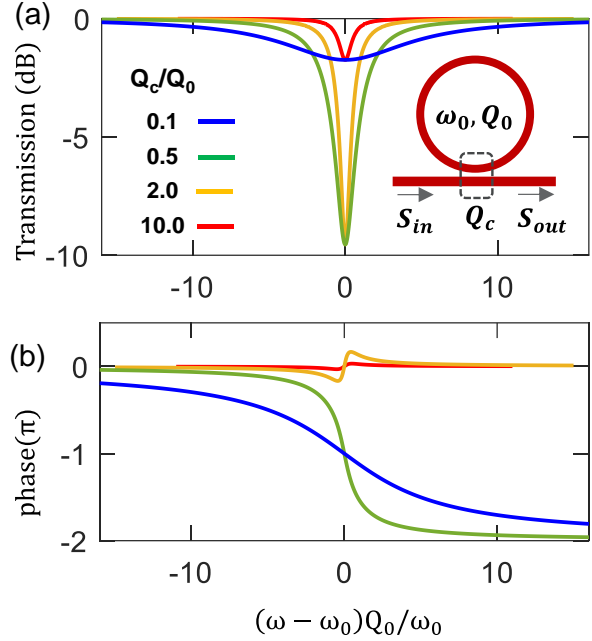

**Fig. 2**: **Ring resonator as a phase shifter**: (a)-(b) Transmission and phase spectrum of a waveguide coupled to a ring resonator for different coupling strengths in the undercoupled ($Q_c/Q_0 >1$) and overcoupled ($Q_c/Q_0 <1$) regimes. The inset in (a) shows the general schematic of the waveguide and resonator with resonance frequency ($\omega_0$), intrinsic Q ($Q_0$) and coupling Q ($Q_c$). A strong overcoupled regime (e.g. $Q_c/Q_0$=0.1, blue plot) provides a large phase change with almost $2\pi$ phase variation when sweeping the laser and resonance frequency relative to each other, and with minimum amplitude distortion of the transmission.

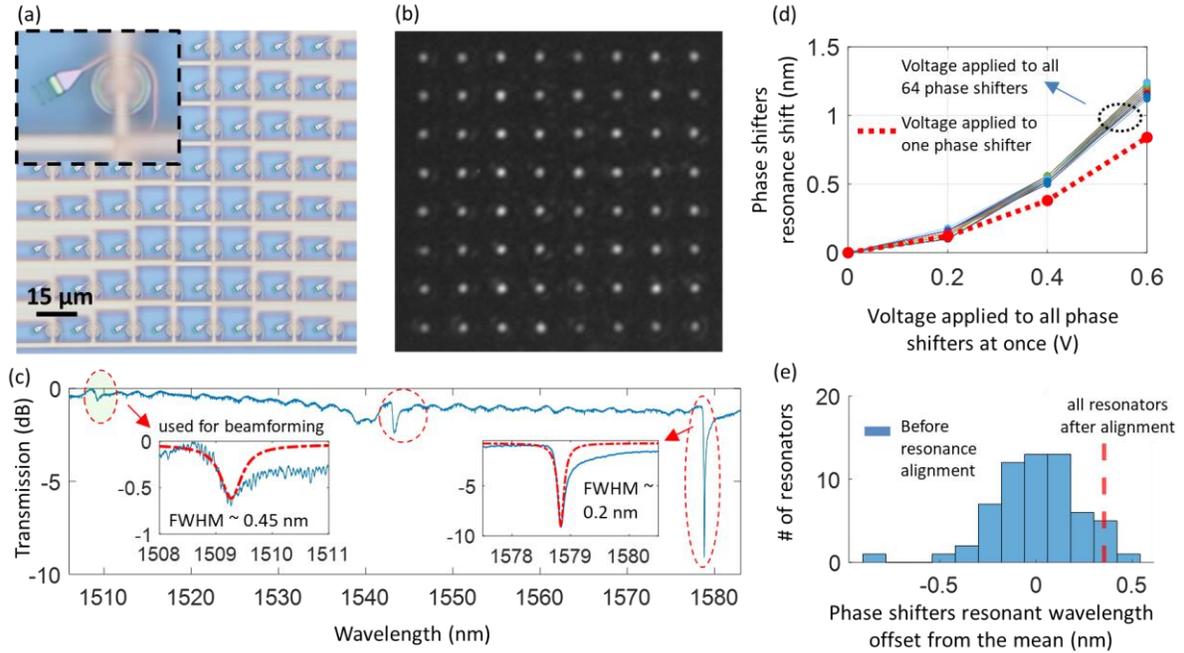

**Fig. 3**: (a) A microscope view of the OPA, with a zoomed view in the inset. (b) The measured near field intensity of the OPA near 1550 nm, showing uniform intensity across the array. (c) A representative phase shifter resonance spectrum, wherein three of the resonances have been highlighted and show different levels of extinction. The right resonance has a larger extinction and narrower linewidth (FWHM~0.2 nm), while the most left resonance is in the strongly over-coupled regime, i.e. shallow (~0.6 dB) extinction, and with a FWHM ~0.45 nm (see the Supplementary Section S2). We use this resonance for far field beam forming and steering. (d) Resonance shift of all 64 phase shifters versus the applied voltage. (e) Statistical variation of untuned resonance wavelengths of all the resonator phase shifters within the 8x8 PIC OPA. This variation is due to fabrication imperfections. The dashed red line is the value all the resonance wavelengths are tuned to and lined up by applying voltage the microheater phase shifters. The resonance shift due to the heat power is ~ 1.1 nm/mW.

## Results

Our PIC-OPA is fabricated in a commercial silicon foundry. We design all the devices in this OPA for a transverse electric (TE) polarization (see Supplementary Section S1). Fig. 3(a) shows a microscope image of the 8x8 OPA device, while Fig. 3(b) shows a near field image of light emitted from the OPA when a laser light (near 1550 nm) is coupled to the chip. As seen from this figure, all of the antennas have relatively uniform intensity.

Figure 3(c) shows the spectrum of a representative resonator phase shifter over a broad spectral range, wherein three resonances have been highlighted. The resonance near 1509 nm shows a strongly overcoupled resonance with a small transmission extinction of only ~0.6 dB and with a resonance linewidth of ~ 450 pm. We use this resonance wavelength for the beamforming and steering. We use the resonances at ~1579 nm for aligning all the resonances because these resonances have higher extinction allowing an easier notch detection. This alignment will automatically align all the resonances of other wavelengths (see the Supplementary Section S3 for further details).

The resonator phase shifters which are controlled by doped Si microheaters are high speed and consume low power. The microheaters have a resistor value of ~ 400 Ω. Figure 3d shows the resonance shift of the phase shifters versus the applied voltage. In one case, we apply voltage to all 64 phase shifters to see the impact of the thermal crosstalk, and in the other case we apply the voltage to only one of the phase shifters. We observe that by applying ~0.6 V to each microheater, we obtain a large 1.1 nm resonance shift much larger than the resonance FWHM (0.45 nm). This demonstrates that at low powers we can have a ~$2\pi$ phase shift when tuning the resonance around its center frequency (see Fig. 2(b)). In addition, these microheater phase shifter show a high-speed frequency response ~330 KHz (see Supplementary Materials) enabling high-speed spatial light modulation and beamforming.

Though the resonator phase shifters are identical in design, the fabrication imperfections cause a variation in their resonance wavelengths resulting in a statistical distribution of resonance wavelength for each of the 64 resonators, as shown in Fig. 3(e). However, these resonance differences are relatively small and can be compensated with a small bias voltage before calibration for far field beam formation, with each resonator being aligned to the position shown by the red dotted line in Fig. 3(e). We expect by using a better fabrication foundry to achieve less resonance wavelength variation.

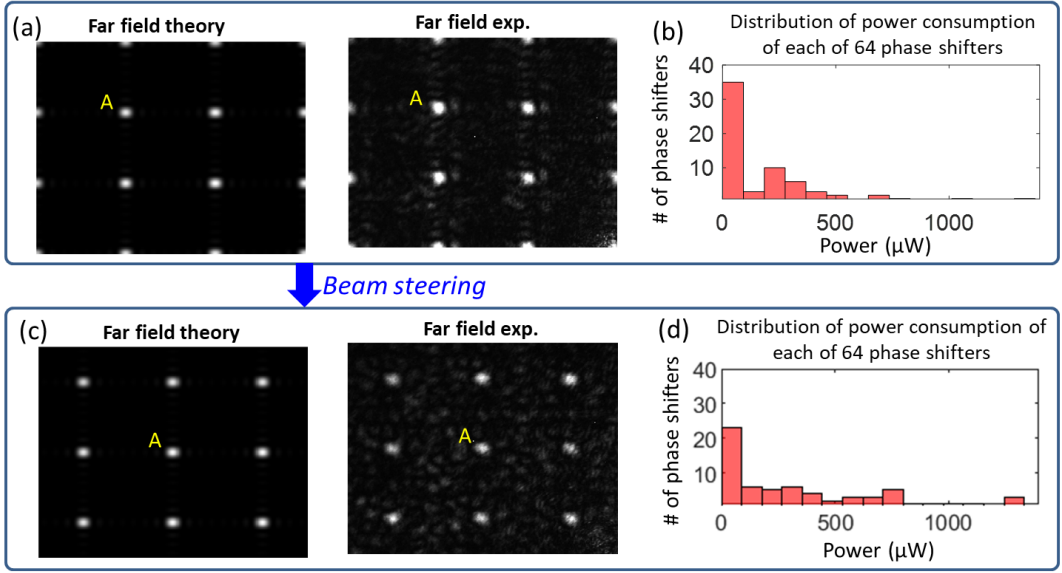

**Fig. 4**: **Far field beam forming and steering.** (a) Simulated and experimental far field for steering angles of $\phi_x = 0$, $\phi_y = 0$. (b) The power consumption range of the phase shifters, for the far field in panel (a). (c)-(d) Repeat of the experiment in panel (a)-(b) but for steering phase (angles) of $\phi_x = \pi$ (2.9º), $\phi_y = \pi$ (2.9º). One of the beam spots has been labeled with letter A to clearly show its displacement in both horizontal (to the right) and vertical (to the down) when comparing panels (a) and (c). For all the beamforming and steering experiments, the average electric power consumption per phase shifter is ~ 300 μW, wherein 250 μW is static for the resonance alignment, and 50 μW is dynamic for beamforming and steering.

To achieve the desired far field beam pattern, we program the phase of each of the 64 microresonator phase-shifters by optimizing the far field beam patterns according to simulated reference patterns (see Supplementary Section S4 for details). Thermal crosstalk of the phase shifters is compensated for in our far field beam optimization without crosstalk measurement.

We achieve beamforming and steering with a low average power of ~300 μW per resonator phase shifter where most of this power consumption (~250 μW) is static (bias), i.e. to align the resonance frequency of the resonators deviated by fabrication imperfections. Figures 4(a) and 4(c) show the simulated far field pattern and the corresponding experimental results for two different beamforming points. We have labeled one of the far field beam spots by letter A in Figs. 4(a) and 4(c) to show where the beam moves. After calibration, the beam can be steered to any desired point in the far field.

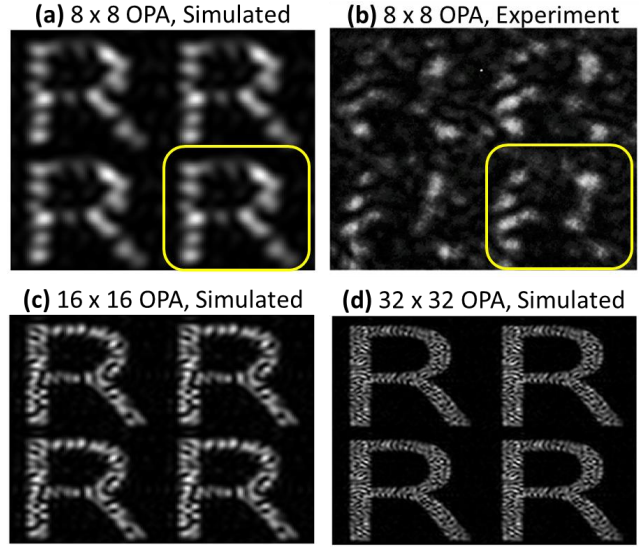

**Fig. 5**: **Far field pattern writing**. (a) Simulated far field of the 8 x 8 OPA for four repetitions of the letter R. (b) Experimental measured data corresponding to simulated data in (a). One of the R letters have been highlighted with yellow rectangle in panels (a) and (b) for clear comparison of theory vs. experiment. The lower image resolution in (a) and (b) is due to the small size OPA (i.e. 8 x8). The panels in (c) and (d) show the simulated far field for a larger array with 16 x 16, and 32 x 32 elements with a higher image resolution.

For the optimizations of the far field beam patterns, we utilize two techniques in series (see also Supplementary Sections S3 and S4). First, we process the phased arrays in the near field using a swept source to extract resonance spectra of the resonator phase shifters. Using a tunable laser, we sweep the laser wavelength while monitoring the radiated nearfield intensity of each antenna via imaging and extract each corresponding phase shifter spectrum. After this measurement and identifying the resonances, we align all the resonance wavelengths by applying voltage to the phase shifters, and find a lookup table of voltages which aligns all the resonances. From this point, we use a genetic algorithm to optimize the far field beam pattern and find lookup tables of voltages for different beam locations. Gradient-based search algorithms and variants are a popular choice for optical phased array optimization and calibration [21-24], though calibration can also be achieved analytically [25]. Figures 4(b) as well as 4(d) show the statistics of the power consumption of each phase shifter within the array for the corresponding far field beams shown in the panels (a) and (c). The average power for each phase shifter is less than 300 µW for all the steering angles within the 0-$\pi$ phase shift, while about ~250 µW is static for the resonance alignment, and ~50 µW is for beam forming.

In large-scale 2D optical phased arrays with many elements, the projection of arbitrary images in the far field is possible by tuning the phases on the antennas. In addition to forming beams for beam steering, we also form the letter "R" in the far field with our OPA device for a proof of concept for arbitrary beam forming applications. Far field pattern formation resolution is limited by number of array elements, but with an 8x8 array, we are able simulate (see Fig. 5(a)) and experimentally match with our OPA device (see Fig. 5(b)) a recognizable "R" in the far field. As simulated far field projections with increasing number of array elements show in Fig. 5(c-d), as the number of array elements within an OPA increases, better resolution of far field patterns is achievable.

## Conclusion

In this work we have demonstrated a silicon photonic 2D OPA as a low power programmable SLM with far field beamforming and steering. Our PIC OPA consists of an 8×8 array with ultracompact and low power resonator phase shifters. With our 8x8 OPA we perform beamforming and 2D beam steering as well as writing patterns in the far field. For these experiments, each phase shifter consumes an average electric power of 300 µW power where most of it is static consumption to align the resonances due to fabrication imperfections. A future improved fabrication in a better foundry lithography node is expected to reduce the static power by at least a factor of two. Though in this work we demonstrate a 2D OPA with periodic pixels, one can place antennas in random locations with an appropriate design algorithm to enable a random sparse array [11].

Larger scale versions of our PIC-OPA SLM device with more array elements can be achieved by implementing electronic integration in a secondary layer below the optical layer. In addition, and for further scalability, one can form a super array composed of our OPA blocks, wherein optical signal is routed between these blocks. Our architecture shows great promise for a new generation of compact and efficient programmable photonic processors by combining the unique properties of planar and free-space optics.

**Conflict of Interests**

M. Soltani, J. Leatham, A. Singh are involved in developing PIC-OPA at Raytheon.

**Acknowledgements**

The authors acknowledge Drs. Stephen Palese, Duane Smith, Alex Latshaw, Mr. Richard Kendrick, and Ms. Charley Fodran, for the fruitful technical discussion on this work. This document does not contain technology or technical data controlled under either the U.S. International Traffic in Arms Regulations or the U.S. Export Administration Regulations.

**Authors Contribution:** All authors contributed to all aspect of this work.

**Data availability**: The data that support the findings of this study are available from the corresponding authors upon reasonable request

# Supplementary Materials:

# Scalable and ultralow power silicon photonic two-dimensional phased array

Michelle Chalupnik[1,2,5], Anshuman Singh[1], James Leatham[3], Marko Loncar[4], Moe Soltani[1]

[1]Raytheon BBN, 10 Moulton Street Cambridge, MA 02138, USA
[2]Department of Physics, Harvard University, Cambridge, MA 02138, USA
[3]Raytheon Intelligence & Space, 2000 El Segundo Dr., El Segundo, CA, USA
[4]John Paulson School of Engineering and Applied Science, Harvard University, Cambridge, MA 02138, USA
[5]Currently with Aliro Quantum, Brighton, MA 02135, USA

*michelle.chalupnik@raytheon.com, mo.soltani@raytheon.com*

**S1. Detailed design parameters of the OPA**

**S2. Detailed discussion on spectrum of the resonator phase shifter**

**S3. Discussion on the resonance alignment of all the resonator phase shifters in the OPA**

**S4. Discussion on the far field beamforming and steering algorithm optimization**

**S5. Phase shifter response time characterization**

## S1. Detailed design parameters of the OPA

Figure S1 shows more details of the PIC OPA we designed in this work. Figure S1a and S1b show the schematic of the OPA and one of its unit cells. In Fig. S1a, evanescent tap couplers are designed properly (see also [1]) and with different lengths to provide equal power per antenna. The Si waveguide width in all of the OPA circuits except at the coupling region to the resonator is 450 nm. As shown in Fig. S1b, for an efficient waveguide-resonator coupling we use a pulley coupling scheme [2, 3]. The simulations show a waveguide width (~340 nm) to closely phase match the waveguide mode to that of the ring resonator mode. The ring resonator has an external radius of 3 micron and a width of 530 nm.

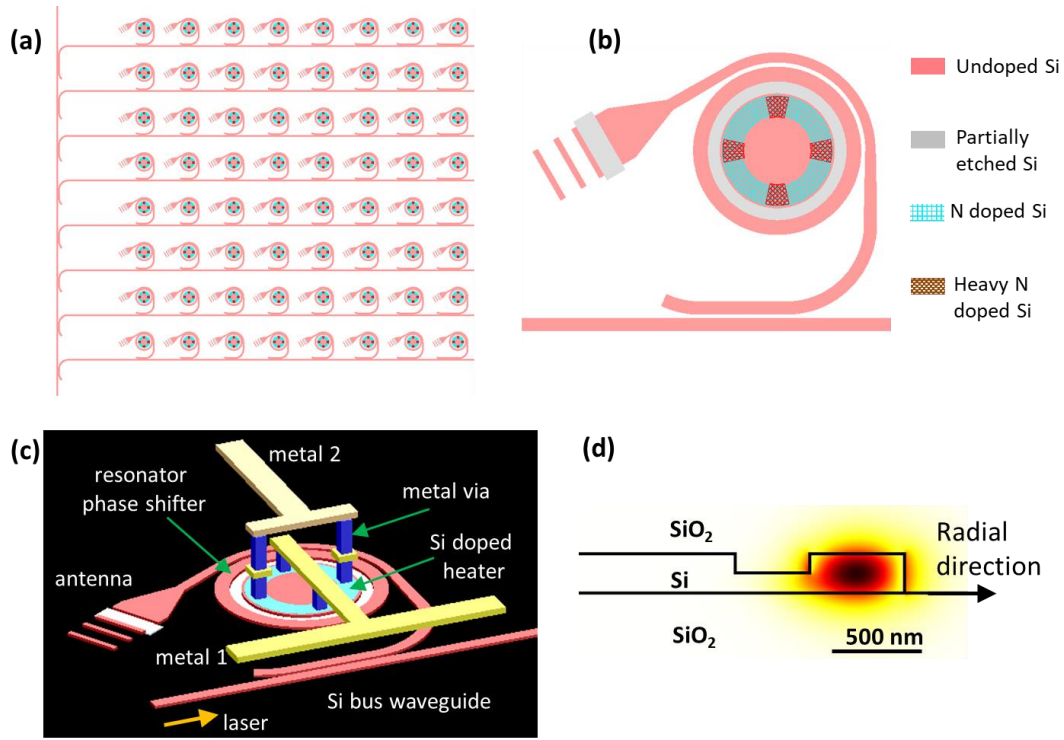

**Fig. S1**: (a) Schematic of the 8x8 OPA which consists of 8 horizontal bus waveguides with equal powers tapped from the coupler coupled to the vertical bus waveguide. On each horizontal bus waveguide there are 8 tap couplers which all equally receive powers to send to their unit cell that includes a phase shifter and an antenna. The metal interconnect is not shown. (b) Schematic of a unit cell. The phase shifter is a ring resonator with a radius of 3 micron and width of 530 nm. The interior of the ring resonator is partially etched with a width of 400 nm. The center of the resonator is a Si disk region where the doped heater and the heavy doped contacts are implemented on the full height silicon (~220 nm). A pulley coupling scheme couples the resonator to a waveguide. The waveguide width as the pulley region is ~ 340 nm, and the waveguide resonator gap is 170 nm. The waveguide output of the phase shifter is connected to an antenna with details described in [2]. (d) The cross section mode profile of the resonator for the transverse electric (TE) polarization.

The center of the resonator is a Si disk region where the perimeter of this disk has four segmented N doped regions and four heavily doped regions to provide four resistive heaters in parallel with the appropriate metal contacts. A partially etched Si region with a width of 400 nm separates the ring resonator from the center Si disk. This partially etched region has two purposes. The partially etched Si region provides confinement for the resonator mode and it prevents the mode leakage to the doped Si region, and hence reduces the optical absorption loss due to the doped region. Our theoretical simulation shows an absorption limited intrinsic Q of ~ 300,000 due to the Si doped region. The partially etched Si region also allows more efficient heat transfer from the Si doped heater to the ring resonator mode.

In the unit cell shown in Fig. S1b, the optical signal after the interaction with the resonator phase shifter is send to an optical antenna with the detailed designs discussed in our earlier work [2]. The radiation efficiency of this antenna is ~55%. Figure S1c is the unit cell with the two metal layers placing four resistive segments in parallel. The total net resistor of the four parallel resistors is ~ 400 Ω. Figure S1d shows the cross section of the simulated electric mode profile the resonator for the TE polarization. The free-spectral-range (FSR) of this resonator mode is ~ 35 nm.

## S2. Detailed discussion on spectrum of the resonator phase shifter

Figures S2a and S2b show the resonance spectrum of two different OPA chips over a broad range of spectra. From these measurements three resonances for the TE mode of the resonator are observable at wavelengths of ~ 1509 nm, 1553.5 nm, and 1579 nm. In each panel, the fit of the resonance spectrum to the theory using coupled mode theory [5] provides the $Q_0$, $Q_c$, and the FWHM.

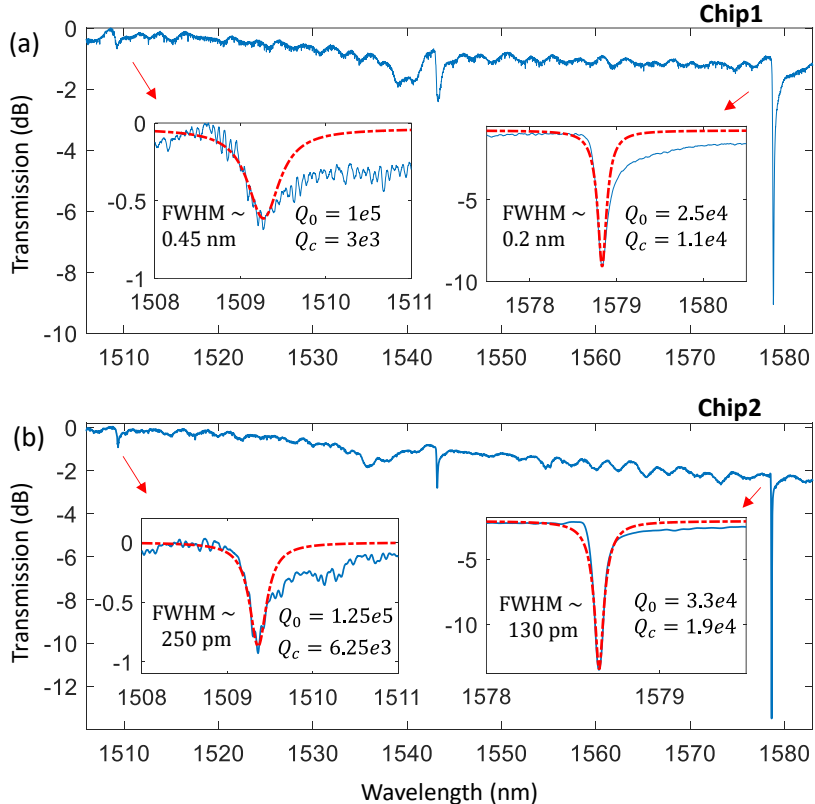

**Fig. S2**: (a) Resonance spectrum of the OPA chip used for the experiments. Three resonances at wavelengths ~1509nm, ~1543.5nm, and ~1579 nm. The inset shows the fit of the experiment to theory to extract the FWHM, the intrinsic Q ($Q_0$), coupling Q ($Q_c$). The resonance at ~1509nm is used for the phase shifter operation. The resonance at ~1579 nm is used for aligning the resonances as it has a deeper extinction making the resonance notch detection and alignment easier. (b) Resonance spectrum for another OPA chip which shows narrower resonance linewidth. In this paper we characterized the Chip 1 for the beamforming and beam steering.

The resonance near 1509 nm is strongly overcoupled, and we use this wavelength region and the resonance for phase shifter operation. The resonance near 1579 nm has large extinction, and we used this resonance wavelength region for aligning all the resonances, as the higher extinction made the notch minimum detection and alignment easier.

For our experiments we used the Chip 1 shown in Fig. S2a for the characterization of the OPA and beamforming and steering. Figure S3a shows the near field intensity image of the OPA, and Fig. S3b shows a representative resonance spectrum and two resonance wavelength regions that we used for the experiments.

## S3. Discussion on the resonance alignment of all the resonator phase shifters in the OPA

To perform calibration of the beam steering lookup tables, we first align all 64 resonances. We measure the near field intensity of each antenna of the OPA over a spectral wavelength range around 1509 nm and 1579 nm to find the resonance spectra of the phase shifter for these two wavelength ranges. Once we have found the resonance spectrum for all 64 phase shifters, we proceed to the resonance alignment. Measuring the notch resonance wavelength of all the resonators in the wavelength region of 1579 nm and applying a proper feedback voltage to the phase shifter we align all the resonances. Aligning these resonance will result in aligned resonances at ~1509 m, which are the resonances we use for phase shifting.

Figure S3a and S3b show the near field intensity of the OPA array and a representative resonance spectrum wherein we have highlighted the two resonances of interest for alignment algorithm (~1579 nm) and beamforming/steering (~1509 nm). In Fig. S3a we have highlighted three rows of the OPA near field intensity, and have shown their corresponding spectrum in Figs. S3c-S3h, after the resonance alignment. As seen from these spectrum figures, the resonances are all aligned, and more importantly, all the resonances at ~1509 nm are strongly overcoupled such that their extinction is buried in the background transmission.

## S4. Discussion on the far field beamforming and steering algorithm optimization

After aligning each resonance, we finish the calibration for beam steering lookup tables. We chose a genetic algorithm to calibrate our OPA to avoid the necessity of crosstalk compensation through measurement. Our genetic algorithm requires a population of sets of lookup table voltages to undergo mutation, combination, and culling, ultimately resulting in an optimized set of voltages which produce an optimized far field beam pattern. In the implementation of the particular genetic algorithm used in this work, we define several hyperparameters, the fraction f1 of the population of sets of voltages which will undergo mutation, the fraction f2 of the population of sets of voltages which will undergo combination, the fraction f3 of the population of sets of voltages which will survive after each iteration, the number of elements per individual which will undergo mutation, the total population size, and the number of repeating periods in the far field considered. First, a fraction f1 of the population changes a set number of their elements to random voltages between the set bounds on voltage. Then, a fraction f2 of the population undergoes combination, where pairs of randomly selected population members randomly switch half of their elements. At the end of the iteration, the fitness of each population member is evaluated, and the top fraction f3 of the remaining population is kept for the next iteration. The fitness is equal to the summed pixel value difference between the simulated reference image and the experimental image.

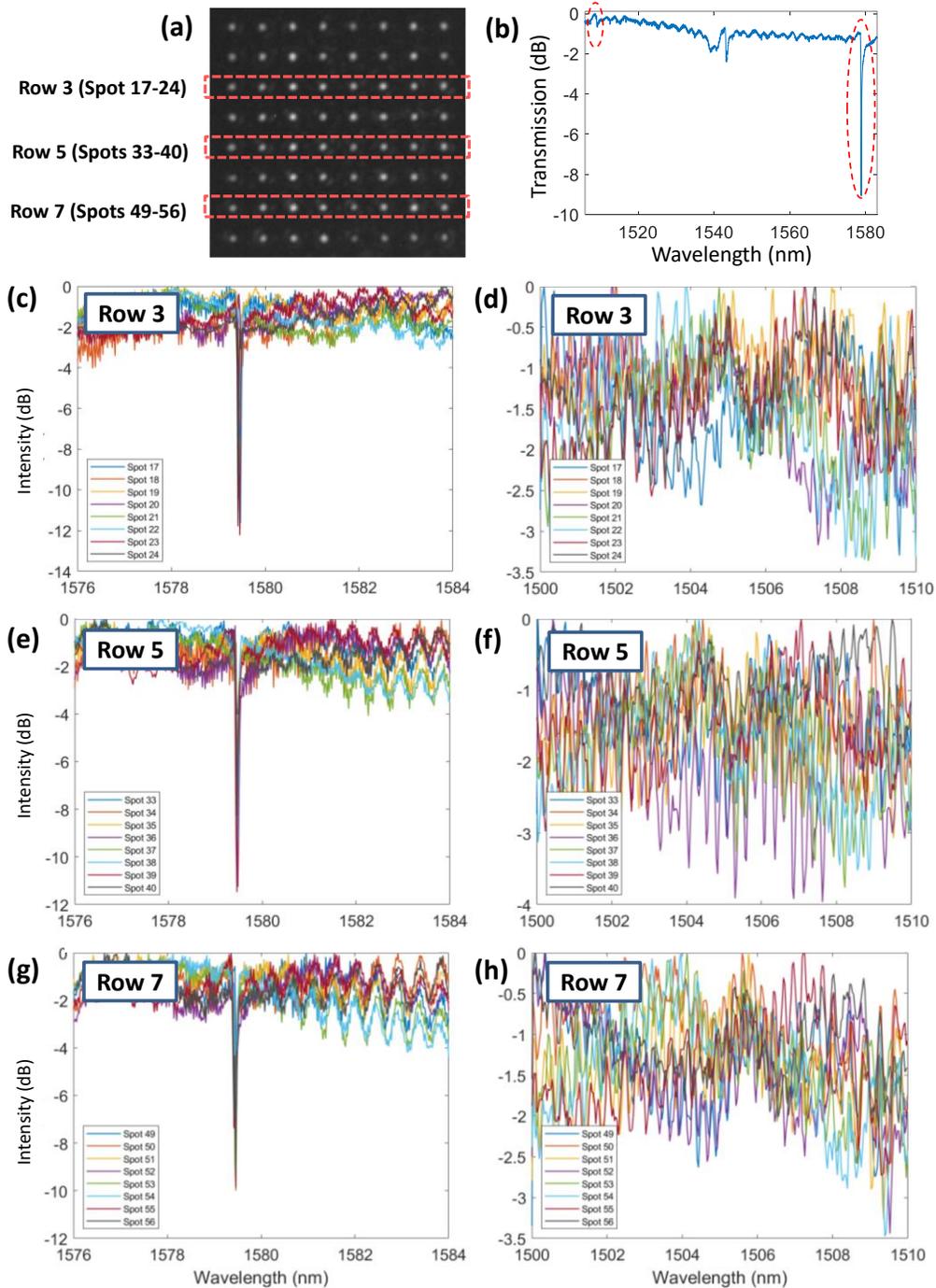

**Fig. S3**: **Aligning of all of the 64 resonators in the 8x8 OPA**. (a) Near field intensity view of the OPA. Three rows have been identified that we will show their resonance alignment data. (b) A representative resonator spectrum. For the resonance alignment we use the resonances near ~1579 nm due to the large extinction allowing easily observable resonance dips. Aligning these resonances will result in aligned resonances near ~1509 nm which we use for beamforming and steering. (c)-(d), (e)-(f), and (g)-h) show the aligned resonances for the rows 3, 5, and 7 in the OPA shown in (a). As seen in the right panels, the resonances near ~1509 nm are all strongly overcoupled such that their extinction are buried in the noise.

## S5. Phase shifter response

To characterize the frequency response of the resonator phase shifter controlled by the microheater, we tune the laser near the resonance, and apply a voltage square wave voltage the microheater with a peak voltage of 0.3 V, and monitor the optical transmission of the resonator phase shifter through its coupling waveguide sent to a high-speed photodetector. Figure S4 shows the measured results where we have laid the square wave voltage (blue) and the output waveguide intensity measured (red) by the photodetector. From the red plot in Fig. S4 and measuring the amplitude decay time we find a response time of ~ 3μs for this microheater phase shifter corresponding to a frequency response of ~ 330 KHz.

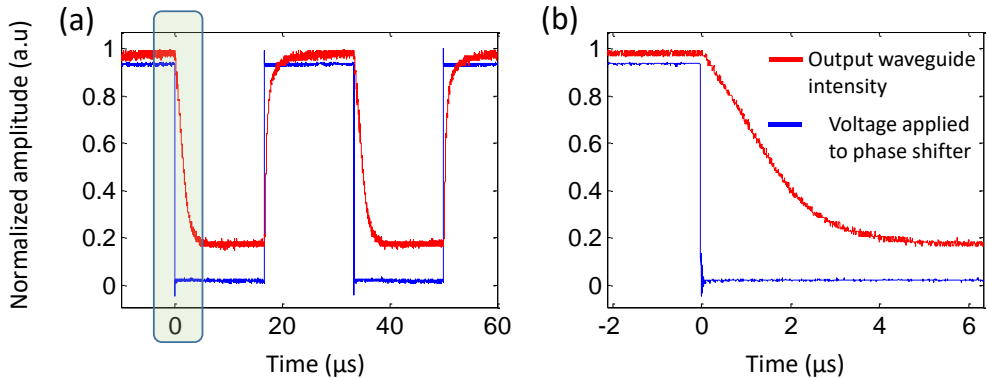

**Fig. S4**: **Microheater resonator phase shifter time response** (a) Plot of the square applied voltage (blue) with an amplitude of 0.3 V to the resonator phase shifter which is laser-tuned to its resonance, and the corresponding variation of the optical intensity because of the resonance shift due to the applied voltage. (b) a zoomed view of the panel (a) in the highlighted green region. From (b) we see a decay time of ~ 3 μs corresponding to a frequency response of ~ 330 KHz.